\documentclass[pdftex,5p,twocolumn]{elsarticle}
\usepackage{txfonts}
\usepackage{natbib}
\bibpunct{(}{)}{;}{a}{}{,}
\usepackage{indentfirst} 
\usepackage{amssymb}
\usepackage{array}
\usepackage{verbatim}

\begin{document}

\begin{frontmatter}

\title{Transport of solids in protoplanetary disks: Comparing meteorites and astrophysical models}

\author[cita]{Emmanuel Jacquet\corref{cor1}}
\ead{ejacquet@cita.utoronto.ca}
\cortext[cor1]{Corresponding author}

\address[cita]{Canadian Institute for Theoretical Astrophysics, University of Toronto, 60 St Georges Street, Toronto, ON M5S 3H8, Canada}

\begin{abstract}
We review models of chondrite component transport in the gaseous protoplanetary disk. Refractory inclusions were likely transported by turbulent diffusion and possible early disk expansion, and required low turbulence for their subsequent preservation in the disk, possibly in a dead zone. Chondrules were produced locally but did not necessarily accrete shortly after formation. Water may have been enhanced in the inner disk because of inward drift of solids from further out, but likely not by more than a factor of a few. Incomplete condensation in chondrites may be due to slow reaction kinetics during temperature decrease. While carbonaceous chondrite compositions might be reproduced in a ``two-component'' picture (Anders 1964), such components would not correspond to simple petrographic constituents, although part of the refractory element fractionations in chondrites may be due to the inward drift of refractory inclusions. Overall, considerations of chondrite component transport alone favor an earlier formation for carbonaceous chondrites relative to their noncarbonaceous counterparts, but independent objections have yet to be resolved. 
 \end{abstract}

\end{frontmatter}

\section{Introduction}
With the accelerating pace of exoplanet detections
, the protoplanetary disk phase of stellar evolution enjoys considerable interest. Thanks to increasing computational power, theorists can test mechanisms for disk transport \citep{Turneretal2014} and planet formation \citep{YoudinKenyon2013}. Observations of present-day protoplanetary disks \citep{WilliamsCieza2011} probe the disk mass, size, structure, chemical species and solids \citep{Nattaetal2007}. However, even with the Atacama Large Millimeter/Submillimeter Array
, it will remain challenging to resolve scales below a few AUs and probe the optically thick midplane of the inner disks where planet formation should occur.  To understand the evolution of solids in disks, we must turn our attention to constraints provided closer to us by our own solar system in the form of primitive meteorites, or \textit{chondrites}. Indeed, chondrites date back to the protoplanetary disk phase of the solar system, 4.57 Ga ago, and with more than 40,000 specimens classified to date, not to mention samples returned from comet Wild 2 \citep[e.g.][]{Zolenskyetal2008Wild2} or asteroid 
Itokawa \citep{Nakamuraetal2011}, they offer a considerable wealth of petrographic, chemical, and isotopic data at all examination scales.

  Yet chondrites arrive in our laboratories without geological context. While orbit determinations consistently assign their parent bodies to the asteroid main belt---with the exception of micrometeorites \citep[e.g.][]{EngrandMaurette1998} and perhaps some carbonaceous chondrites \citep{Gounelleetal2008} possibly derived from further out---, exactly where and when they originally accreted is largely unknown. Still, chondrites exhibit considerable compositional variations, and space and time were obviously important dimensions behind them. In fact, each individual chondrite is a mixture of components (chondrules, refractory inclusions, etc.) formed in different locations, epochs and environments in the disk \citep{BrearleyJones1998, Krotetal2009}. This is evidence for considerable transport in the disk. In order to place the meteoritical record in context, the relevant transport processes have to be understood, and as such meteorites are sensors of the dynamics of protoplanetary disks.

  Our purpose here is to review transport mechanisms of chondrite components before accretion. An earlier review on particle-gas dynamics was given by \citet{CuzziWeidenschilling2006} and \citet{Boss2012} reviewed transport and mixing from the perspective of isotopic heterogeneity. The formation \textit{per se} of chondrite components is essentially beyond our scope but the reader may be referred to recent reviews by \citet{Krotetal2009} and \citet{Aleon2010}. \citet{Wood2005} and \citet{Chambers2006} proposed syntheses on the origin of chondrite types from cosmochemical and astrophysical viewpoints, respectively. Here, the discussion will be organized around meteoritical constraints as follows: In section \ref{Background}, we provide background on chondrites and the basic physics of the protoplanetary disk before embarking in section \ref{Transport} on an examination of transport constraints from specific chondrite components. We then review the interpretation of fractionation trends exhibited by chondrites as wholes (section \ref{Fractionation}) in light of which we will discuss the chronological and/or spatial ordering of chondrite groups (section \ref{Chondrites in space and time}).

\section{Background}
\label{Background}
\subsection{Chondrites: a brief presentation}
  Chondrites are assemblages of various mm- and sub-mm-sized solids native to the protoplanetary disk. Oldest among them are the \textit{refractory inclusions} \citep{MacPherson2005,Krotetal2004}, further divided in calcium-aluminum-rich inclusions (CAI) and (less refractory) amoeboid olivine aggregates (AOA), which presumably originated by high-temperature gas-solid condensation, 4568 Ma ago \citep{BouvierWadhwa2010,Connellyetal2012,Kitaetal2013}, although many have since experienced melting. More abundant than those are \textit{chondrules}, silicate spheroids 1-4 Ma younger than refractory inclusions \citep{KitaUshikubo2012,Connellyetal2012}, likely formed by melting of isotopically and chemically diverse precursor material. The nature of the melting events remains however elusive, with ``nebular'' (e.g. shock waves) and ``planetary'' (e.g. collisions) environments still being considered \citep{Boss1996,Deschetal2012}. \textit{Metal} and \textit{sulfide} grains also occur, either inside or outside chondrules \citep{Campbelletal2005}. All these components are set in a fine-grained \textit{matrix}, a complex mixture of presolar grains, nebular condensates and/or smoke condensed during chondrule-forming events \citep{Brearley1996}. 

  While all chondrites roughly exhibit solar abundances for nonvolatile elements \citep{PalmeJones2005}, with CI chondrites providing the best match, they are petrographically, chemically and isotopically diverse, and 14 discrete \textit{chemical groups}, each believed to represent a distinct parent body (or a family of similar ones), have hitherto been recognized. To first order, one may partition these groups in two \textit{super-clans} \citep{Kallemeynetal1996,Warren2011b}, namely the \textit{carbonaceous chondrites} (with the CI, CM, CO, CV, CK, CR, CB, CH groups), and the \textit{non-carbonaceous chondrites}, which comprise the enstatite (EH, EL), ordinary (H, L, LL) and Rumuruti (R) chondrites. Carbonaceous chondrites are more ``primitive'' in the sense that they have a higher abundance of refractory inclusions and matrix, a solar Mg/Si ratio, and an $^{16}$O-rich oxygen isotopic composition closer to that of the Sun \citep{McKeeganetal2011}. Non-carbonaceous chondrites, though poorer in refractory elements, are more depleted in volatile elements, have subsolar Mg/Si ratios and a more terrestrial isotopic composition for many elements \citep[e.g.][]{Trinquieretal2009}. What these differences mean and how some may relate to the transport of chondrite components is one of the main focuses of this review.

\subsection{Dynamics of the early solar system}
\label{Disks}

The exact structure of our protoplanetary disk remains very conjectural. If we mentally add gas to a smoothed version of the current planetary system to restore solar abundances, we obtain a density profile known as the ``Minimum Mass Solar Nebula'' (MMSN; \citet{Hayashi1981}) with an integrated mass $\sim$0.01 M$_\odot$ (1 M$_\odot$ $\equiv$ 1 solar mass). While this agrees with disk masses estimated for most T Tauri stars \citep{WilliamsCieza2011}, it may be one order of magnitude below the original disk mass at the cessation of infall from the parent molecular cloud \citep[e.g.][]{YangCiesla2012}. The MMSN, though a useful reference, ignores the extensive redistribution and losses of gas and solids occurring in disks which funnel gas onto the central stars at observed rates of $10^{-8\pm 1}\:\mathrm{M_\odot/a}$ \citep{WilliamsCieza2011}.

  What drives the evolution of gas disks? Since the molecular viscosity is far too small to account for the $\sim$1-10 Ma lifetime of protoplanetary disks \citep{WilliamsCieza2011}, disk theorists generally rely on turbulence which, in a \textit{rough, large-scale} sense, may mimic the effects of an enhanced viscosity
\begin{equation}
\label{nu}
\nu=\alpha \frac{c_s^2}{\Omega},
\end{equation}
with $c_s$ the (isothermal) sound speed, $\Omega$ the Keplerian angular velocity and $\alpha$ the dimensionless ``turbulence parameter'' (see e.g. \citet{BalbusPapaloizou1999}), for which values around $10^{-2}$ are inferred from observations \citep{Armitage2011}. The exact source of this turbulence is still contentious. As yet, the leading candidates are \textit{gravitational instabilities} \citep{Durisenetal2007} and the \textit{magneto-rotational instability} (MRI; \citet{BalbusHawley1998}). While the former would be important in the earliest epochs where the disk is massive enough, the latter essentially only requires the gas ionization fraction to be above a small threshold, and may operate at all times. However, this threshold may not be attained over a considerable range of heliocentric distances, yielding a \textit{dead zone} of low turbulence, unless other instabilities are at work (see \citet{Turneretal2014}). Eventually, after the disk mass has significantly dropped, photoevaporation due to the central star (and/or close neighbours) should completely clear the gas \citep{Armitage2011}.

  Except in late stages of the disk where photophoresis may become important  \citep{WurmKrauss2006}, the dynamics of small solids are primarily dictated by \textit{gas drag}. The stopping time, for a spherical particle of radius $a$ smaller than the molecular mean free path, is \citep{Weidenschilling1977}:
\begin{equation}
\label{tstop}
\tau=\sqrt{\frac{\pi}{8}}\frac{\rho_sa}{\rho c_s},
\end{equation}
with $\rho_s$ and $\rho$ the solid and gas densities, respectively. For instance, at 3 AU in a MMSN, a 0.3 mm radius chondrule has $\tau = 2$ h at the midplane, much shorter than the orbital period. Hence, to zeroth order, chondrite components should follow the gas, but they cannot have \textit{exactly} the same velocity because they do not ``feel'' the pressure gradient acceleration experienced by it. There is thus a systematic drift velocity of the solids relative to the gas in the direction of larger pressures, that is, toward both the midplane and the Sun. The radial drift is given by:
\begin{eqnarray}
v_{\rm drift,R}=\frac{\tau}{\rho}\frac{\partial P}{\partial R}=0.004\:\mathrm{m/s}\frac{\partial\mathrm{ln}P}{\partial\mathrm{ln}R}\left(\frac{\rho_sa}{1\:\rm kg/m^2}\right)\left(\frac{10^{-6}\:\rm kg/m^3}{\rho}\right)\nonumber\\
\left(\frac{1\:\rm AU}{R}\right)\left(\frac{T}{300\:K}\right)^{1/2}
\end{eqnarray}
with $P$ and $T$ the gas pressure and temperature and $R$ the heliocentric distance. While this drift is generally smaller than the turbulent velocity fluctuations of the gas ($\sim\sqrt{\alpha}c_s$), it may become important in the long run as these average out. A measure of this importance is the ``gas-grain decoupling parameter'' \citep{Cuzzietal1996,Jacquetetal2012S}:
\begin{eqnarray}
\label{S}
S & \equiv & \frac{\Omega\tau}{\alpha}
\\
 &=& 0.1\left(\frac{\rho_sa}{1\:\rm kg/m^2}\right)\left(\frac{10^{-8}\:\rm M_\odot/a}{\dot{M}}\right)\left(\frac{R}{1\:\rm AU}\right)^{3/2}\left(\frac{T}{300\:\rm K}\right),\nonumber
\end{eqnarray}
where the last equality is for a steady disk of mass accretion rate $\dot{M}$. For $S\ll 1$, the particles are tightly coupled to the gas, while for $S\gtrsim 1$, they settle to the midplane (with a concentration factor $\sim\sqrt{S}$) and drift radially sunward faster than the gas.

  Over time, grains collide and coagulate, as evidenced by detection, in protoplanetary disks, of mm-sized solids \citep{Nattaetal2007}, much larger than typical interstellar grains ($\sim 0.1\:\mu$m). Growth to meter size is frustrated by bouncing/fragmentation at high collision speeds and by increased radial drift which would remove them from the disk within centuries \citep{Weidenschilling1977,Braueretal2007,Birnstieletal2010}. Settling to the midplane might help self-gravity of the solids to intervene but is limited by gas turbulence, so that other mechanisms such as turbulent concentration or streaming instabilities may have to bridge the gap \citep{CuzziWeidenschilling2006,YoudinKenyon2013}. Here, we will mostly restrict attention to sub-mm/cm-sized bodies since chondrite composition was established at the agglomeration of such components.

\begin{figure}
\resizebox{\hsize}{!}{
\includegraphics{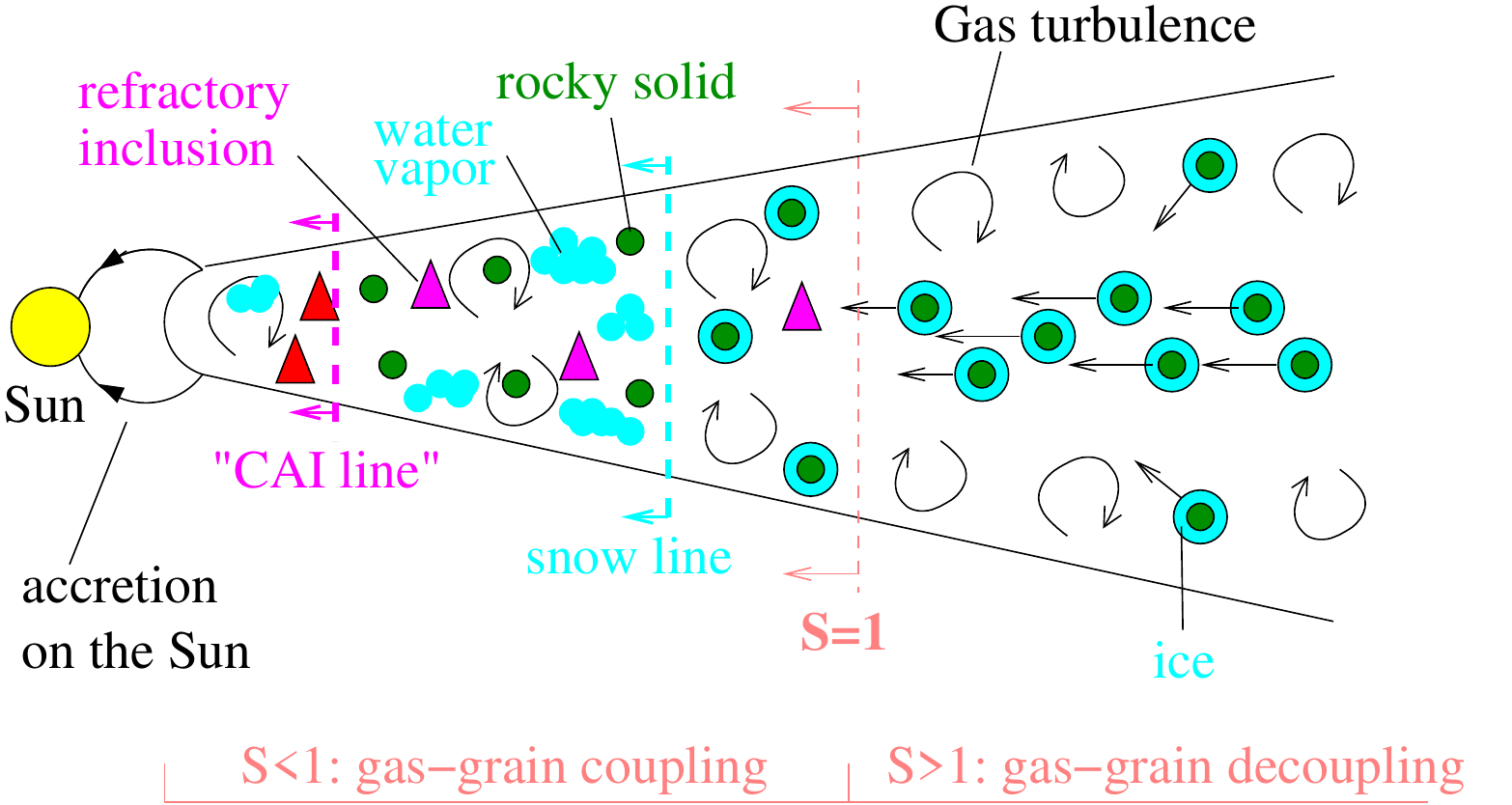}
}
\caption{Sketch of dynamics in the early solar system, with refractory inclusions (pink), rocky solids (green) which may be thought of as chondrules or chondrule precursors, and water (cyan). Turbulent motions of the gas are indicated (with the understanding that their average incur a net flow toward the Sun). Beyond the $S=1$ line (which depends on the particle size), solids decouple from the gas, settling to the midplane and drifting toward the Sun faster than the gas. Lines corresponding to CAI (for the early stages) and water condensation are shown as well, with an indication of their secular inward displacement as the disk evolves.
}
\label{schema}
\end{figure}

\section{Transport of chondrite components}
\label{Transport}

With these fundamentals in mind, we now turn to constraints provided by specific chondrite components, and the astrophysical processes which may satisfy them.

\subsection{Refractory inclusions}
\label{CAI}

The oldest solids of the solar system, the refractory inclusions, are estimated to have formed at $\sim$1400-1800 K \citep{Grossman2010}. Such high temperatures, presumably due to small-scale dissipation of turbulence, did not obtain at heliocentric distances $>$ 1 AU for more than a few $10^5$ years after disk formation \citep[e.g.][]{YangCiesla2012}. This is consistent with the old age of CAIs and AOAs; yet to account for their presence in chondrites and even in comets \citep[e.g.][]{Simonetal2008Inti}, outward transport is in order. It would also account for the abundance of crystalline silicates in comets \citep[e.g.][]{Zolenskyetal2008Wild2}.

  \citet{Shuetal1996} proposed that solids processed at $R<0.1$ AU were entrained by stellar winds and fell back onto the disk further out \citep{Hu2010}. \citet{Deschetal2010} however criticized the role of this ``X-wind'' in processing chondrite components e.g. regarding the very survival of solids (see also \citet{Cuzzietal2005} on the formation timescales of Wark-Lovering rims) so that one may have to seek transport within the disk itself.

  In a turbulent disk, velocity fluctuations may send high-temperature material outward \citep[e.g.][]{Gail2001,BockeleeMorvanetal2002,Cuzzietal2003,Boss2004,Ciesla2010,HughesArmitage2010}. The envisioned random-walk paths may explain complex thermal histories recorded by some refractory inclusions \citep{Ciesla2011radial,Bossetal2012}. The efficiency of this turbulent diffusion against the mean inward flows depends sensitively on the Schmidt number Sc$_R\equiv\nu/D_R$, with $D_R$ the turbulent diffusivity, although many studies have simply equated it to 1. Sc$_R<1$ seems required for efficient outward diffusion \citep{ClarkePringle1988,PavlyuchenkovDullemond2007,HughesArmitage2010}. While MRI-driven turbulence would likely not satisfy this requirement \citep{Johansenetal2006}, this is expected from hydrodynamical turbulence  \citep{Prinn1990}, but save for a few laboratory experiments, of uncertain relevance to protoplanetary disks \citep[e.g.][]{Launder1976,Lathropetal1992}, empirical evidence is largely wanting.

  Turbulent diffusion may have been supplemented by \textit{outward} advection flows early in the disk evolution. Indeed, the disk may have been initially compact ($\lesssim$ 10 AU in radius), and its ensuing expansion would have begun in the condensation region of refractory inclusions \citep{Jacquetetal2011a,YangCiesla2012}. Outward transport would then have been efficient for the earliest generation of CAIs, hence perhaps their narrow \textit{observed} age range \citep{Ciesla2010,YangCiesla2012}. 

  Outward flows have been proposed to persist around the disk midplane, even if the vertically integrated flow is inward, and ease the outward transport of inner disk material at later stages \citep{Ciesla2007,HughesArmitage2010}. This so-called meridional circulation arises when turbulence is modeled as a viscosity in a very literal sense \citep[e.g.][]{TakeuchiLin2002}. However, while the gross properties of turbulent disks may be obtained with this ansatz, there is no first principle reason that the resulting two-dimensional flow structure should also hold true, and in fact meridional circulation has not been observed in numerical simulations of MRI-driven turbulence by \citet{Fromangetal2011} or \citet{Flocketal2011}. At any rate, \citet{Jacquet2013} showed that even if meridional circulation existed, it would not, because of the inward flows in the upper layers, make a significant difference in terms of net radial transport compared to 1D models, given current uncertainties in turbulence parameters.

  Not only were refractory inclusions transported in the disk, they were preserved there quite efficiently, for the CAI fraction in some carbonaceous chondrites is comparable to what \textit{in situ} condensation out of a solar gas would have produced ($\sim$6 \%; \citet{Grossman2010}). However, the inward gas flows and grain-gas radial drift have long been expected to remove them from the chondrite-forming regions within a few $10^5$ years---especially for the large type B CAIs in CV chondrites---, even though the drift slows down closer to the Sun \citep{Laibeetal2012}. Indeed turbulent diffusion calculations by \citet{Cuzzietal2003} and \citet{Ciesla2010} underpredicted CAI abundances by 1-2 orders of magnitude---unless the ``CAI factory'' was enriched in condensible matter, which, if not very carbon-rich, would however yield too oxidizing conditions \citep{Jacquetetal2011a}. In simulations starting with compact disks, however, \citet{YangCiesla2012} achieved retention of refractory inclusions for $>$2 Ma, presumably because many of them were sent far from the Sun (10-100 AU), their disk remained quite massive ($\gtrsim 0.1\:\rm M_\odot$) even after a few Ma, and $\alpha$ was relatively low ($10^{-3}$; see also equation 8 of \citet{Jacquetetal2011a}). In fact, low turbulence levels $\alpha\lesssim 10^{-4}$ could \textit{alone} account for the preservation of refractory inclusions, assuming outward transport was accomplished somehow before, and low turbulence is exactly what is generically expected from the dead zone picture \citep{Jacquetetal2011a}. Indeed, the dead zone, which would have emerged \textit{after} an initially turbulent phase conducive to extensive transport, would slow down gas accretion toward the Sun, and by forcing gas to accumulate there, would also reduce the stopping time and thence the drift of refractory inclusions.

  Importantly, efficient diffusion as expected in the early disk would rapidly homogenize any short-lived radionuclide like $^{26}$Al \citep[e.g.][]{Bossetal2012} and hence validate its use as a chronometer. Same would not hold, however, if such isotopes were injected into the disk after the formation of a dead zone. 

\subsection{Chondrules}
\label{Chondrules}

  The low turbulence levels invoked above for the preservation of refractory inclusions would as well account for the few-Ma age range of chondrules measured in single meteorites \citep{KitaUshikubo2012,Connellyetal2012}, for chondrules and refractory inclusions have comparable sizes. However, \citet{AlexanderEbel2012} argued that turbulent mixing would homogenize chondrule populations over chondrite-forming regions within a few $10^5$ years, at variance with the distinctive chondrule populations of the different chemical groups \citep{Jones2012}. They thus suggested that the Al-Mg ages---although broadly corroborated by Pb-Pb dating---were perturbed (but see \citet{KitaUshikubo2012}) and that the data are consistent with chondrule formation immediately preceding chondrite accretion. From an astrophysical standpoint, this may be a premature conclusion, though. The diffusion length after a time $t$ is
\begin{equation}
\sqrt{2D_R t}
=1\:\mathrm{AU}\left(\frac{t}{1\:\rm Ma}\right)^{1/2}\mathrm{Sc}_R^{-1/2}\left(\frac{\alpha}{10^{-4}}\right)^{1/2}\left(\frac{T}{300\:\rm K}\right)^{1/2}\left(\frac{R}{1\:\rm AU}\right)^{3/4}.
\end{equation}
So whether this exceeds the separation between different chondrule-forming regions depends, among other things, on the exact values of $\alpha$ and on where the chondrule- and chondrite-forming regions actually were in the disk. These locales may have been quite distinct from the present-day position of chondrite parent bodies, in the asteroid main belt, especially if their orbits were significantly reshuffled e.g. during a ``Grand Tack'' \citep{Walshetal2011}. Moreover, this calculation ignores the barrier that gas drag-induced drift may have posed to outward mixing (if $S>1$). In fact, if mixing had been as efficient as to homogenize the chondrite-forming region in $\lesssim 1$ Ma timescales, bulk chemical fractionations observed across chondrite groups would be difficult to understand (see section \ref{Fractionation}).

Another constraint that chondrule transport must satisfy is \textit{chondrule-matrix complementarity} \citep{Hussetal2005}. This is the observation, for carbonaceous chondrites, that while the bulk rocks have solar Mg/Si ratios (or other interelement ratios, see \citet{HezelPalme2010}), this does not hold for chondrules or matrix taken individually. Complementarity, if confirmed (but see  \citet{Zandaetal2012}), requires that chondrules and matrix be genetically related (unlike, e.g., a X-wind scenario), but also that chondrules and dust from a given chondrule-forming region did not drift apart until accretion. While a particular chondrule and a particular dust grain would quickly separate barring immediate accretion, this would not be true of the populations of chondrules and dust grains \textit{as wholes} which would remain spatially indistinguishable for some time due to turbulence. In fact, in the regime $S<1$ (for chondrules), \citet{Jacquetetal2012S} showed that this overlap would continue over their whole drift timescale, so that accretion at any time would yield complementarity. Complementarity would not be compromised by mixing between products of several chondrule-forming events provided transport from each of these sources was likewise unbiased as to the chondrule/dust ratio \citep{Cuzzietal2005,Jacquetetal2012S}. Chondrule transport in the disk is thus still compatible with observations, although a link between \textit{chondrule} and \textit{chondrite} formation cannot be ruled out.

\subsection{Metal and sulfide grains}
\label{Metal}

Chondrites, and especially non-carbonaceous ones---plus the very metal-rich CHs and CBs, although their genesis likely was very anomalous \citep[e.g.][]{Krotetal2005}---, have undergone metal/silicate fractionation prior to accretion \citep{LarimerWasson1988,Wood2005}. While early workers invoked some separation of metal grains directly condensed from the hot solar nebula, e.g. because of ferromagnetically enhanced coagulation (\citet{HarrisTozer1967}; see also a review by \citet{Kerridge1977}), there is little evidence of \textit{isolated} pristine nebular metal condensates in meteorites, although such grains are found \textit{enclosed} in refractory inclusions \citep{Weisbergetal2004,Schwanderetal2013}. Actually, chondrite 
metal grains mostly seem to be byproducts of chondrule-forming events \citep{Campbelletal2005}. Metal/silicate fractionation may have arisen locally by aerodynamic sorting \citep[e.g.][]{Zandaetal2006}, e.g. because of differential radial and/or vertical drift, or turbulent concentration. Indeed, in both ordinary \citep{Kuebleretal1999,NettlesMcSween2006} and enstatite \citep{Schneideretal1998} chondrites, metal and sulfide grains have a somewhat lower $\rho_sa$ than chondrules on average, but are closest to aerodynamic equivalence with them for their most Fe-rich varieties (H and EH, respectively) which have the smaller chondrules. Therefore, metal/sulfide grains and small chondrules could have been segregated together. Alternatively, metal/silicate fractionation might reflect varying contributions of debris of differentiated planetesimals predating chondrule formation (e.g. \citet{SandersScott2012}; but see \citet{FischerGoeddeetal2010}).

\subsection{Matrix grains}
The grains of chondritic matrices are typically sub-$\mu$m-sized \citep{PontoppidanBrearley2010}, likely too small to show any decoupling relative to the nebular gas prior to agglomeration as fluffy aggregates or fine-grained rims around chondrules \citep[e.g.][]{MetzlerBischoff1996}. For $\alpha=10^{-3}$, surface densities below 1 kg/m$^2$ (2 orders of magnitude lower than the MMSN at 30 AU) would be required to see any effect on radial drift (i.e. $S\gtrsim 1$). It is then surprising that silicate and sulfide grains in chondritic porous interplanetary dust particles are aerodynamically equivalent \citep{Wozniakiewiczetal2012}; if non-coincidental, it could indicate very low densities in the outer disk when these (likely comet-derived) objects formed. 

  While no differential drift of dust grains is expected anyway in the inner disk, presolar grains show some variations across chondrite groups, e.g. the proportions of type X SiC grains \citep{Zinner2003} or the carriers of $^{54}$Cr anomalies \citep[e.g.][]{Trinquieretal2009}. These may be due to non uniform injection or thermal processing of these grains \citep{Trinquieretal2009} and their persistence (at a few tenths of the anomalies of refractory inclusions)
 suggest limited turbulence levels, as similarly inferred above for refractory inclusions and chondrules. As the high-temperature events which produced the latter would have destroyed presolar grains, their very survival indicates that these events were localized, allowing subsequent mixing between processed and unprocessed matter.

\subsection{Water}
Water makes up about half of condensable matter in a solar mix \citep{Lodders2003}. While chondrites that survive atmospheric entry are mostly dry, hydrated silicates, mostly found in carbonaceous chondrites, testify to aqueous alteration on their parent body \citep{Brearley2003}. Also, water may have been partly responsible for the high oxygen fugacities recorded by many chondrules \citep[e.g.][]{Schraderetal2013}, which require 10-1000-fold enhancements over solar abundances. Water was likely $^{16}$O-poor \citep{Sakamotoetal2007}, and possibly responsible for the variations of the oxygen isotopic composition of the inner solar system, from $^{16}$O-rich signatures of CAIs to $^{16}$O-poor, ``planetary'' ones (but see \citet{Krotetal2010}; see also \citet{Yurimotoetal2008} for an overview of oxygen isotopic data).

  In protoplanetary disks, water condenses as ice beyond the ``snow line'' ($\sim$170 K). Ice and intermingled silicates would drift inward and enrich the inner disk inside the snow line \citep{StepinskiValaegas1997,CuzziZahnle2004}. In the popular CO self-shielding scenario, whether in the parental molecular cloud \citep{YurimotoKuramoto2004} or in the disk \citep{Lyonsetal2009}, as $^{16}$O-poor water may be most efficiently produced and/or preserved at large heliocentric distances, this would account for its addition to inner solar system material, although detailed calculations of O isotopic evolution in disks have yet to be published (it remains in particular to be seen whether the existence of \textit{both} $^{16}$O-rich and -poor reservoirs already during CAI formation as recorded by some reversely zoned melilite grains \citep{Parketal2012} can be reproduced). Because of the finite supply of water in the outer disk (and/or the ``bouncing barrier'' to grain growth which would limit drift), the enrichment would be limited to a factor of a few \citep{CieslaCuzzi2006,HughesArmitage2012}, insufficient to account for FeO contents in chondrules. Settling to the midplane might further enhance the (dust $\pm$ ice)/gas ratio to the desired levels, depending on the enhancement due to radial drift, but would require very low turbulence \citep{CuzziWeidenschilling2006}. Turbulent concentration is yet another possibility \citep{Cuzzietal2001}.

  For efficient ice accretion beyond the snow line \citep[e.g.][]{StevensonLunine1988}, diffusion may later \textit{deplete} the inner disk in water, perhaps accounting for the (reduced) enstatite chondrites \citep{Paseketal2005}, although replenishment from further out would limit this to a factor of a few \citep{CieslaCuzzi2006}. 

  Another important constraint on water is the D/H ratio which, for carbonaceous chondrites appears systematically lower than most comets \citep{Alexanderetal2012}. This may require efficient outward diffusion (i.e. low Schmidt number) of D-poor water from the warm inner disk (\citet{JacquetRobert2013}; see also \citet{Yangetal2013}), consistent with the requirement of efficient outward transport of high-temperature minerals (\citet{BockeleeMorvanetal2002}; see section \ref{CAI}).

\section{Fractionation trends in chondrites}
\label{Fractionation}

We have investigated above the constraints given by individual petrographic components of chondrites on their transport in the protoplanetary disk. On a more integrated perspective, such redistribution of material may have caused the compositional variations exhibited by the different chemical groups of chondrites. We have already mentioned metal/silicate fractionation (section \ref{Metal}); here, we focus on lithophile element fractionations and their possible dynamical interpretations.

  With respect to solar abundances, the most striking pattern exhibited by bulk chondrite chemistry (except CIs) is the depletion in volatile elements, increasing with decreasing nominal condensation temperature \citep{Palmeetal1988}. How did this \textit{incomplete condensation} come about? \citet{Yin2005} suggested that it was inherited from the interstellar medium, but isotope systems involving elements of different volatilities (e.g. Rb-Sr) yield whole-rock isochrons consistent with the age of the solar system \citep{PalmeJones2005}, indicating, along with the very existence of undepleted CI chondrites native to the solar system, that the depletion arose in the disk itself. \citet{Cassen1996} reproduced some of the elemental trends by assuming that the chondrite parent bodies started to form while the disk was hot and massive, but this is inconsistent with more recent evidence that chondrites accreted after a few Ma \citep{Ciesla2008}. Timescale considerations also exclude the suggestions by \citet{WassonChou1974} of gas-solid separation by settling, radial drift (both of which require $S>1$ which would obtain late in the disk history \citep{Jacquetetal2012S}), or gas photoevaporation. The last plausible alternative may then be a slowing of reaction kinetics upon decrease of temperature---which are anyway required to explain the preservation of CAIs in the first place \citep{Ciesla2008}. Whether the temperature changes witnessed by individual condensates (over 10-1000 years in simulations by \citet{Bossetal2012} and \citet{Taillifetetal2013}) would be sufficiently rapid to incur such an effect has yet to be investigated.     

  Whatever process caused incomplete condensation, it did not operate to the same degree in all regions and epochs of the disk, and undoubtedly, there has been mixing by diffusion and differential drift between these different reservoirs. For example, CM chondrites are enriched in refractory lithophile elements but moderately volatile elements exhibit a plateau at about half the CI chondritic value, suggesting a 50 \% admixture of CI-like material to an otherwise smoothly volatile-depleted material \citep{Cassen1996}. \citet{Anders1964} proposed that the composition of chondrites resulted from varying proportions of an unfractionated CI chondritic and a high-temperature component. \citet{Zandaetal2006} recently developed this \textit{two-component model} by identifying these components with petrographic constituents such as CAIs, chondrules and matrix, whose proportions would have varied independently accross the different chondrite groups (this may be called the ``strong'' two-component model). It is however questionable whether these petrographic components were dynamically independent from each other, in particular for carbonaceous chondrites. In the regime $S<1$, which would hold for the first few Ma, there would be indeed little decoupling between these. Observational evidence for coherence between chondrite components is provided by (i) matrix-chondrule complementarity (see section \ref{Chondrules}) and (ii) the \textit{subsolar} Al/Si ratios of \textit{CAI-subtracted} carbonaceous chondrites \citep{Hezeletal2008}, contrary to a simple picture of CAI addition to a CI chondritic material, suggesting a genetic link between at least some of the CAIs and their host chondrite \citep{Jacquetetal2012S}.  Also, the distinctiveness of chondrules in different chondrite groups \citep{Jones2012} excludes that a single chondrule population was distributed throughout the disk. Thus, while carbonaceous chondrite bulk compositions might conceivably be modeled in a simple two-component picture \citep{Zandaetal2012}, with higher high-temperature fractions presumably representing earlier times and/or shorter heliocentric distances, such chemical components would likely have no straightforward petrographic manifestation. 

  Non-carbonaceous chondrites, to which we now turn attention, are depleted in refractory lithophile elements relative to CIs, but this trend does not actually simply complement the enrichment exhibited by carbonaceous chondrites, as it is accompanied by a decrease in the Mg/Si ratios (roughly uniformly solar for carbonaceous chondrites; \citealt{LarimerWasson1988refractory}). Another process must be at play. \citet{LarimerWasson1988refractory} proposed a loss of a refractory olivine-rich material, possibly AOAs \citep[see also][]{Ruzickaetal2012AOA}. This could be accomplished by inward drift to the Sun (in the regime $S>1$) provided that this component was present in grains systematically coarser than the other \citep{Jacquetetal2012S}---at least before chondrule formation.  \citet{Hutchison2002} proposed instead the addition of low Mg/Si material to CI composition to account for non-carbonaceous chondrite composition, which may be implemented in a X-wind model but would be subject to the drawbacks of such scenarios \citep{Deschetal2010}. In the case of enstatite chondrites, which have the lowest Mg/Si ratios of chondrite groups, \citet{Lehneretal2013} proposed that sulfidation of silicates may have led to evaporative loss of Mg. The concentration of chondrules, e.g. due to preferential settling \citep{Jacquetetal2012S} or turbulent concentration \citep{Cuzzietal2001}, might explain the volatile-depleted composition of non-carbonaceous chondrites relative to their carbonaceous counterparts.

\section{Chondrites in space and time}
\label{Chondrites in space and time}

In this final section, we would like to return to our original question---how the different chondrite groups may be ordered in space and time in the early protoplanetary disk.

  It is widely assumed that chondrite groups represent different heliocentric distances of formation, with enstatite chondrites closest to the Sun, followed by ordinary, Rumuruti and carbonaceous chondrites \citep[e.g.][]{RubinWasson1995,Wood2005,Warren2011b}. Certainly, spectroscopic observations---and the sample return mission to S(IV) asteroid Itokawa \citep{Nakamuraetal2011}---suggest that enstatite, ordinary and carbonaceous chondrites are associated with E, S, and C-type asteroids, respectively, and these do exhibit this radial sequence \citep{Burbineetal2008}, although with wide overlap (e.g. \citet{Usuietal2013} find that most of the \textit{large} E-type asteroids actually lie in the \textit{middle} of the asteroid belt).

  \textit{Ab initio} rationalization of this trend as a purely spatial effect is however problematic. It is e.g. no longer possible to ascribe the implied increase in oxidation state with heliocentric distance to a temperature decrease as in the classic ``hot solar nebula'' picture, not only because high temperatures would not have prevailed long, but also because ferroan olivine in chondrites is difficult to ascribe to nebular condensation \citep{Grossmanetal2012}; and in fact, \textit{chondrules} in carbonaceous chondrites are \textit{more reduced} than their ordinary chondrite counterparts. The nonmonotonic trend in isotopic ratios of oxygen or other elements \citep{Warren2011b} is also difficult to ascribe to episodic infall \citep{RubinWasson1995} as infall would long have ceased. 

  Could time of formation have then played a role? We have seen in the previous section that carbonaceous chondrites were enriched in refractory elements (in particular in CAIs) compared to non-carbonaceous chondrites. Regardless of the details of the fractionation mechanisms, if, from the above, they did not form closer to the Sun, it seems unavoidable that they formed earlier, as proposed by \citet{Cuzzietal2003} (specifically for CV chondrites with their large type B CAIs, similar to \citet{Wood2005}) and \citet{Chambers2006}. \citet{Jacquetetal2012S} showed that the retention of CAIs required, along with other properties, that $S<1$ for carbonaceous chondrites, and which would also plead in favor of an earlier epoch, as $S$ tends to increase with time and heliocentric distance (see equation (\ref{S})). The fact that noncarbonaceous chondrite parent bodies seem on average closer to the Sun may be due to inward drift which tended to increasingly concentrate solids in the inner regions. Then, the $^{16}$O-poorer composition of non-carbonaceous chondrites may be ascribed to a later, more advanced stage of influx of $^{16}$O-poor water from the outer disk if the self-shielding picture holds.

  While modelling of chondrite component transport thus suggests that carbonaceous chondrites accreted earlier than non-carbonaceous chondrites, other points of view are allowed by other lines of evidence. One is that chondrules in CO and LL chondrites exhibit a similar range of Al-Mg ages ($\sim 1-3$ Ma after CAIs, with younger ages in CRs and EHs \citep{KitaUshikubo2012,Guanetal2006}). While not strictly contradicting a difference in chondrite \textit{accretion} time, this could indeed suggest that time was not an important factor. Also, non-carbonaceous chondrites have generally been more thermally metamorphosed on their parent body than carbonaceous chondrites \citep{Hussetal2006}. If the heating is ascribed to $^{26}$Al decay, which should decrease over time, this would on the contrary suggest that carbonaceous chondrites accreted \textit{later} than non-carbonaceous chondrites (see \citet{GrimmMcSween1993}), unless the higher water content of the former or some difference in the structure or size of the parent bodies was responsible (\citet{Chambers2006}; see also \citet{ElkinsTantonetal2011}).

  It thus appears that there are cogent arguments for the three possible chronological orderings of carbonaceous and noncarbonaceous chondrites (with the former either older, younger or contemporaneous with the latter). Obviously, however, two of these reasonings have to give, but it may still be dicey to decide which with any authority. Resolution of this critical issue in the interpretation of the meteoritical record will await further advances on the transport of chondrite components, as reviewed here, but also on their formation models as well as the thermal and collisional evolution of the chondrite parent bodies.

\section*{Acknowledgments}
Reviews by Prof. Fred Ciesla, Jeffrey Cuzzi and Alexander Krot were greatly appreciated. I am grateful to the ``Meteoritics and Solar System History'' reading group at the University of Toronto for wide-ranging discussions. This review is dedicated to the memory of Guillaume Barlet (1985-2014), a friend and tireless promoter of synergies between astronomers and cosmochemists.

\bibliographystyle{natbib}
\bibliography{bibliography}

\end{document}